\documentclass[pra, twocolumn, floatfix]{revtex4}
\usepackage{graphicx}
\usepackage{amsmath, amsfonts, amssymb, bm}
\usepackage{braket}
\usepackage{color}
\usepackage[utf8]{inputenc}
\begin{document}
\title{Two-center dielectronic recombination in slow atomic collisions.}
\author{ A. Jacob, C. M\"uller and A. B. Voitkiv}
\affiliation{ Institut f\"ur Theoretische Physik I, 
Heinrich Heine Universit\"at D\"usseldorf, 
\\ Universit\"atsstr. 1, 40225 D\"usseldorf, Germany }
\date{\today}
\begin{abstract} 

A free electron can form a bound state with an atomic center $A$ 
upon photo emission (radiative recombination). 
In the presence of a neighboring atom $B$, such 
a bound state can, under certain conditions, be also 
formed via resonant transfer of energy to $B$, 
with its subsequent relaxation through radiative decay 
(two-center dielectronic recombination). 
This two-center process is very efficient  
in the 'static' case where $A$ and $B$ form a weakly bound system,  
dominating over single-center radiative recombination up to 
internuclear distances as large as several nanometers. 
Here we study its dynamic variant in which recombination occurs when 
a beam of species $A$ collides with a gas of atoms $B$ and show that,  
even though the average distance between $A$ and $B$ in collisions 
is orders of magnitude larger than the typical size of a bound system, 
the two-center recombination can still outperform  
the single-center radiative recombination. 

\end{abstract}

\maketitle

\section{ Introduction }

Processes of recombination of free electrons with atomic or molecular ions are of general interest and relevance to various scientific disciplines \cite {Hahn} - \cite{Beiersdorfer}.   
In case of single atomic centers there are three basic recombination processes. First, the electron can be captured into a bound state by emitting a photon (radiative recombination); this process is 
the time-inverse of photoionization. Second, for certain energies of the incident electron, the recombination can proceed resonantly via formation of an autoionizing state (time-reversed Auger decay) which then 
stabilizes to a bound state through spontaneous radiative decay. 
This process is especially important for low-charged ions. 
Third, an electron can be captured by an ion transferring excess energy 
to another free electron (three-body recombination); this process 
becomes very efficient in high-density plasma, especially when the energy transfer is small. 

When an atom is not isolated in space but close to another atom, 
recombination of a free electron with one of them can, under certain conditions, proceed -- due to two-center electron-electron interaction -- via resonant energy transfer to the other atom which afterwards stabilizes via spontaneous radiative decay \cite{2CDR}, \cite{2CDR_1}. This process, termed two-center dielectronic recombination (2CDR), is rather similar to the 'standard' dielectronic recombination on a single center but, in contrast to the former, relies on the interaction between electrons of different centers. The 2CDR can also be viewed as a kind of three-body recombination in which an assisting free electron is replaced by an electron bound 
in a heavy atomic particle whose internal structure 
plays in this process a crucial role.    
   
It is worth mentioning that the coupling of  
electronic structures at two spatially well separated 
atomic centers  
by long-range electromagnetic interactions can lead  
to a variety of interesting phenomena. 
For example, interatomic electron-electron 
correlations are responsible for 
the population inversion in a He-Ne laser, 
and the energy transfer in quantum optical ensembles \cite{Ficek} 
or cold Rydberg gases \cite{Weidemueller}. They also play an important role 
in biological systems as F\"orster resonances between chromophores 
\cite{Jares}. 
Another interesting realization of two-center electron-electron coupling  
is represented by a process in which the electronic excitation energy 
of one of the atoms cannot be quickly released through a forbidden 
(single-center) Auger decay and is instead transferred to the partner 
atom resulting in its ionization (inter-atomic coulombic decay). 
Stimulated by detailed theoretical 
predictions \cite{ICD}, this process 
has been observed in recent years in various systems such as van der 
Waals clusters \cite{clusters}, rare gas dimers \cite{dimers}, 
and water molecules \cite{ICDexpH2O}.  
In the process of the so called interatomic coulombic electron capture (ICEC), an electron is captured by one atomic center transferring the excess energy to a neigboring atom that results in its ionization 
\cite{Gokhberg}. 

Interatomic electron-electron correlations also drive 
the process of resonant two-center photo ionization (2CPI) \cite{2CPI} 
in which ionization of a van der Waals dimer  
occurs via resonant photo absorption by one of its atoms 
with subsequent transfer of excitation energy 
via two-center electron correlations to another atom 
leading to its ionization. 
This two-center ionization channel can be remarkably effective 
strongly dominating over the usual single-center photo ionization.    
It was experimentally observed in \cite{frank-group} and 
\cite{hergenhahn}.  

It is known \cite{2CDR}, \cite{2CDR_1}, \cite{2CPI} that interatomic electron-electron correlations can greatly enhance recombination and ionization processes in a 'static' situation in which 
two atomic centers constitute a (weakly) bound system. 
The strength of the two-center correlations rapidly decreases with 
increasing the size of the system. Nevertheless, 
it has recently been shown \cite{2CPI_1} that 2CPI can strongly 
dominate single-center photo ionization also in collisions, 
even though the average interatomic distance in collisions  
exceeds by orders of magnitude the typical size 
of the corresponding bound system. 

\begin{figure}[h!] 
\begin{center}
\includegraphics[width=0.48\textwidth]{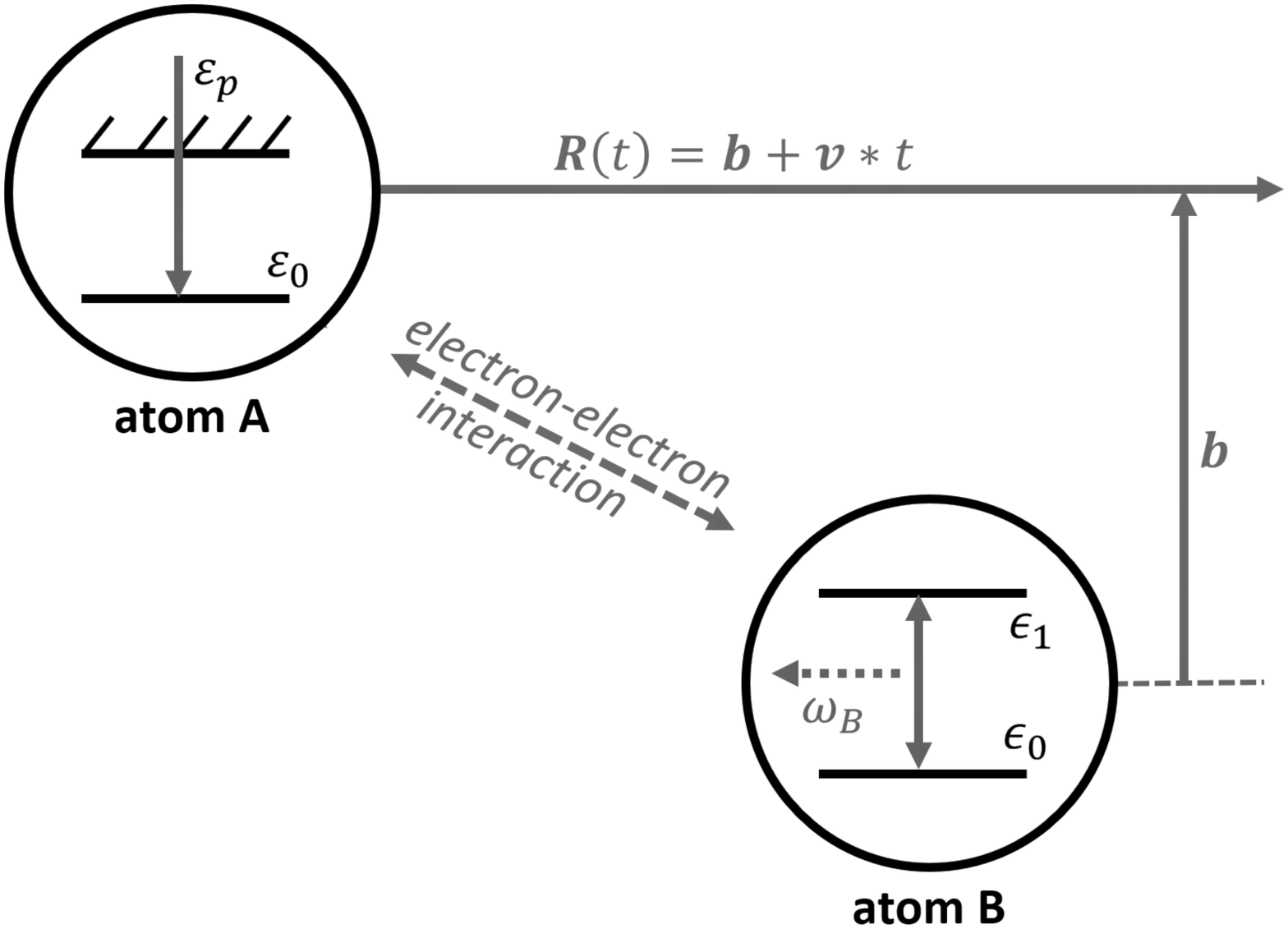}
\end{center}
\caption{Scheme of two-center dielectronic recombination in atomic collisions.}
\label{figure1}
\end{figure}

In this paper, a dynamic variant of two-center dielectronic recombination, 
which occures in atomic collisions (see Fig. \ref{figure1}), is studied. 
We show that although, compared to collisional 2CPI, 
collisional 2CDR turns out to be  much less efficient, it still can outperform single-center radiative recombination (RR). 

The paper is organized as follows. Section II is devoted to considering two-center dielectronic recombination in slow collisions of two atomic centers and to a derivation of formulas for the rates of this process. Besides, in this section we also very briefly discuss 
(single-center) radiative recombination and the collisional version of the inter-atomic coulombic electron capture (ICEC). Section III contains numerical results and discussion and the main conclusions 
are summarized in Section IV.     

Atomic units (a.u.) are used throughout unless otherwise stated.

\section{ Theoretical consideration }

\subsection{ Two-center dielectronic recombination }

Suppose that free electrons and a beam of 
atomic centers $A$ (represented by e.g. ions or atoms) 
move in a (relatively dense) gas of atoms $B$. If the energy release 
in the process of e$^-$ $+$ $A$ recombination   
is close to an excitation energy of a dipole-allowed transition 
in atom $B$, then the recombination can proceed  
by transferring -- via the (long-range) two-center electron-electron interaction -- the energy excess to atom $B$ which, as a result,  
undergoes a transition into an excited state. Afterwards, 
atom $B$ radiatively decays to its initial (ground) state. 

Such recombination relies on the energy transfer 
resonant to a transition in $B$. 
However, the relative motion of $A$ and $B$ 
leads to uncertainty in electron transition energies,  
effectively broadening them. Therefore, the efficiency 
of this recombination channel is expected 
to be restricted to low-velocity collisions. 

Let us now turn to the consideration of recombination 
of a free electron with center $A$ occurring 
in a slow collision between $A$ and atom $B$. 
In order to prevent the resonant nature of the two-center dielectronic recombination from a complete elimination by the relative motion of $A$ and $B$ the collision velocity is supposed to be much less than $1$ a.u. 
($1$ a.u. $ = 2.18 \times 10^8$ cm/s). 

Even though the collision velocity is low, we shall assume that one can still use the semi-classical approximation in which the relative motion of the nuclei of $A$ and $B$ is treated classically. According to the applicability conditions of the semi-classical approximation (which are discussed 
e.g. in \cite{McDowell}) this can be safely done up to impact energies 
as low as $\sim 1$ eV/u. 
Moreover, only (very) distant collisions will be considered 
in which the interaction between $A$ and $B$ is rather weak 
and their relative motion can be approximated by straight lines. 

Let us choose a reference frame in which $B$ is at rest 
and take the position of its nucleus as the origin. 
In this frame $A$ moves along a classical straight-line trajectory ${\bm R} (t) = {\bm b} + {\bm v} t$, where ${\bm b} = (b_x, b_y, 0)$ is the impact parameter and ${\bm v} = (0,0,v)$ the collision velocity. 

The recombination process is described by the Schr\"odinger equation 
\begin{eqnarray}
i \frac{\partial \Psi(t)}{\partial t} = \hat{H} \Psi(t). 
\label{schr-eq}
\end{eqnarray}
Here the total Hamiltonian $\hat{H}$ reads
\begin{eqnarray}
\hat{H}(t) = \hat{H}_A + \hat{H}_B + \hat{H}_{\gamma} + \hat{V}_{AB} + \hat{V}_\gamma, 
\label{total_hamiltonian}
\end{eqnarray}
where $\hat{H}_A$ is the Hamiltonian of the subsystem 
consisting of $A$ and an electron (initially incident and finally 
bound with $A$), $\hat{H}_B$ is the Hamiltonian of the free (non-interacting) atom $B$ 
and 
\begin{eqnarray}
{\hat V}_{AB} = \frac{ {\bm r}\cdot {\bm \xi} }{ R^3 (t) }
 - \frac{ 3({\bm r}\cdot{\bm R} (t) )({\bm \xi}\cdot{\bm R} (t) )}{ R^5 (t) }
\label{2c-ee-int}
\end{eqnarray}
the dipole-dipole interaction between ($A$ $+$ $e^-$) and $B$. 
In Eq.(\ref{2c-ee-int}) the coordinate $\bm r$ refers to the electron recombining with $A$ and is given with respect to the nucleus of $A$ whereas the coordinate ${\bm \xi}$ refers to the electron bound 
in $B$ (and is given with respect to the nucleus of $B$). 
We note that the 'electrostatic' approximation  
(\ref{2c-ee-int}) for the inter-atomic interaction can be used 
if the distance $R$ is not too large: $R \ll c/\omega_{tr}$, where 
$ c $ is the speed of light and $\omega_{tr}$ the frequency 
of the virtual photon transmitting the interaction 
(see e.g. \cite{2CDR}, \cite{2CPI}, \cite{add-ref-2}). 

Further, in (\ref{total_hamiltonian}) $\hat{H}_{\gamma}$ 
is the Hamiltonian of the (quantized) radiation field and 
\begin{eqnarray}
\hat{V}_\gamma = \frac{1}{c} \hat{{\bm A}}({ \bm \xi},t) \cdot \hat{{\bm p}}_{\bm \xi} + \frac{1}{2 c^2} \hat{{\bm A}}^2({ \bm \xi},t) 
\label{V_gamma}
\end{eqnarray}
the interaction of $B$ with this field. 
Here, $ \hat{{\bm p}}_{\bm \xi}$ is the momentum operator 
for the electron bound in atom $B$ and 
\begin{eqnarray}
\hat{{\bm A}}({ \bm \xi},t) = \sqrt{ \frac{2\pi c^2}{V_{ph}\omega_k} } {\bm e}_{ {\bm k} \lambda}
 \left[ \hat{a}_{{\bm k}\lambda} e^{i ( {\bm k} \cdot {\bm \xi} - \omega_k t) } + h.c. \right] 
\label{vect-potential}
\end{eqnarray} 
is the vector potential of the radiation field, 
where $\bm k$ is the wave vector, ${\bm e}_{ {\bm k} \lambda}$ ($\lambda = 1,2$) are the unit polarisation vectors 
(${\bm e}_{ {\bm k} 1} \cdot {\bm e}_{ {\bm k} 2} = 0$, 
${\bm e}_{ {\bm k} \lambda} \cdot {\bm k} = 0$), 
$\omega_k = c k$ is the frequency 
and $V_{ph}$ the normalization volume for the field. 
In what follows the interaction 
$\hat{V}_\gamma$ will be treated in the dipole approximation, 
i.e. ${\bm k} \cdot {\bm \xi} \approx 0$. 

The initial ($\Psi_{{\bm p}0}$), intermediate ($\Psi_{10}$) and final ($\Psi_{00}$) states of the total system -- "($A$ + $e^-$) + $B$ + radiation field" -- are given by 
\begin{eqnarray}
\Psi_{{\bm p}0} ({\bm \xi}, {\bm \rho}, t) &=& \phi_{\bm p} ({\bm \rho - \bm R} (t) ) e^{-i \varepsilon_p t} \alpha ({\bm \rho}, t) \chi_0 ( \bm \xi  ) e^{-i \epsilon_0 t} \nonumber \\
& & \quad \times \ket{0_{{\bm k} \lambda}} \nonumber \\
\Psi_{01} ({\bm \xi}, {\bm \rho}, t) &=& \phi_0 ( {\bm \rho - \bm R} (t) ) e^{-i \varepsilon_0 t} \alpha ({\bm \rho}, t) \chi_1 ( {\bm \xi} ) e^{-i \epsilon_1 t} \nonumber \\
& & \quad \times \ket{0_{{\bm k} \lambda}} \nonumber \\
\Psi_{00} ({\bm \xi}, {\bm \rho}, t) &=& \phi_0 ( {\bm \rho - \bm R} (t) ) e^{-i \varepsilon_0 t} \alpha ({\bm \rho}, t) \chi_0 ({\bm \xi} ) e^{-i \epsilon_0 t} \nonumber \\
& & \quad \times \ket{1_{{\bm k} \lambda}} .\nonumber \\
\end{eqnarray}
Here $\phi_{\bm p}$ is the state of the electron incident on $A$ 
with an asymptotic momentum ${\bm p}$ (as is seen in the rest frame of $A$), $\phi_0$ ($\chi_0$) is the ground state of subsystem (e$^-$+$A$) 
(atom $B$) with an energy $\varepsilon_0$ ($\epsilon_0$), $\chi_1$ the excited state of $B$ with an energy $\epsilon_1$, ${\bm \rho} = {\bm r + \bm R}(t)$ and $\alpha ({\bm \rho}, t) = e^{i {\bm v} \cdot ({\bm \rho - \bm R} (t) )} e^{-i \frac{ v^2}{2}}$ is the translational factor. 
Finally, $\ket{0_{{\bm k} \lambda}}$ ($\ket{1_{{\bm k} \lambda}} $) represents the state of the radiation field before (after) the spontaneous radiative decay in $B$. 

Using the second order of time-dependent perturbation theory and taking into account the selection rules for dipole allowed transitions in atom $B$ ($\Delta l = 1$ and $\Delta m = 0, \pm 1$) for the orbital ($l$) and magnetic ($m$) quantum numbers we obtain that the transition amplitude 
for collisional two-center dielectronic recombination is given by 
\begin{eqnarray}
\mathcal{S}_{2C}^{DR} = \sum_{\Delta m= -1}^{1} \mathcal{S}_{2C}^{DR, \Delta m}. 
\label{total-ampl}
\end{eqnarray}
Here, 
\begin{eqnarray}
\mathcal{S}_{2C}^{DR, \Delta m} = \frac{1}{i^2} \int_{-\infty}^{\infty} dt ~ \mathcal{M}_2^{\Delta m} (t) \int_{-\infty}^{t} dt' ~ \mathcal{M}^{\Delta m}_1(t') ~, 
\label{part-ampl}
\end{eqnarray}
where $\mathcal{M}^{\Delta m}_1 (t') = \bra{ \Psi_{01}} \hat{V}_{AB} \ket{\Psi_{{\bm p} 0}}$ and $\mathcal{M}^{\Delta m}_2 (t) = \bra{ \Psi_{00}} \hat{V}_{\gamma} \ket{\Psi_{01}}$. Integrating in (\ref{part-ampl}) by parts results in 
\begin{eqnarray}
\mathcal{S}_{2C}^{DR, \Delta m} = \sqrt{ \frac{2 \pi}{V_{ph} \omega_k} } \frac{\mathcal{W}_{01}^{B, \Delta m}}{ \frac{\Gamma_r^B}{2} + i \delta} \int_{-\infty}^{\infty} dt ~ \mathcal{M}_1^{\Delta m} (t) e^{-i \delta t}, 
\label{S_2C^DR}
\end{eqnarray} 
where $\mathcal{W}_{01}^{B, \Delta m} = \bra{ \chi_0 ({\bm \xi}) } {\bm e}_{{\bm k} \lambda} \hat{{\bm p}}_{\bm \xi} \ket{ \chi_1 ({\bm \xi}) }$, $\Gamma_r^B$ is the width of the excited state $\chi_1$ due to its spontaneous radiative decay and $\delta = \epsilon_1 - \epsilon_0 - \omega_k $.

Performing the time integration in \eqref{S_2C^DR}, we arrive at
\begin{eqnarray}
\mathcal{S}_{2C}^{DR, \Delta m} &=& \sqrt{ \frac{2^3 \pi}{V_{ph} \omega_k} } \frac{ \lvert \Delta \rvert }{b v^3} \frac{\mathcal{W}_{01}^{B, \Delta m}}{ \frac{\Gamma_r^B}{2} + i \delta} \times \nonumber \\ 
& & \bigg\{ v K_1(\eta) \mathcal{W}_{01{\bm p}0}^{\Delta m} ({\bm \xi}_{\bot} \cdot {\bm r}_{\bot } ) \nonumber \\
 & & - ~ \frac{\lvert \Delta \rvert}{b} K_2(\eta) \mathcal{W}_{01{\bm p}0}^{\Delta m} ( ({\bm \xi}_{\bot} \cdot {\bm b})({\bm r}_{\bot} \cdot {\bm b}) ) \nonumber \\
& & + ~ b \rvert \Delta \lvert K_0(\eta) \mathcal{W}_{01{\bm p}0}^{\Delta m} ( \xi_z r_z ) \nonumber \\
& & + ~ i \Delta K_1(\eta) \times \nonumber \\
& & \quad ~ \mathcal{W}_{01{\bm p}0}^{\Delta m} ( ({\bm \xi}_{\bot} \cdot {\bm b}) r_z + ({\bm r}_{\bot} \cdot {\bm b}) \xi_z ) \bigg\}, 
\label{S_2C_DR}
\end{eqnarray}
where $\eta = \lvert \Delta \rvert \frac{b}{v}$, $\Delta = \varepsilon_p - \varepsilon_0 - \omega_k$, $\mathcal{W}_{01{\bm p}0}^{\Delta m} (x) = \bra{ \phi_0 ({\bm r}) \chi_1 ({\bm \xi})} x \ket{ \phi_{\bm p} ({\bm r}) \chi_0 ({\bm \xi}) }$ ($x \in \mathbb{R}$) and $K_n$ ($n=0,1,2$) are the modified Bessel functions of the second kind \cite{IStegun}. ${\bm r}_{\bot}$ (${\bm \xi}_{\bot}$) is the transverse part of the coordinate ${\bm r}$ ( ${\bm \xi}$ ), which is perpendicular to the collision velocity ${\bm v}$.

The spectra of emitted photons can be calculated from the following quantity 
\begin{eqnarray}
\frac{d^3 \sigma_{2C}^{DR}}{d {\bm k}^3} = 
\frac{ V_{ph} }{ (2 \pi)^3 } \sum_{\lambda} 
\int_{b_{min}}^{\infty} d b ~b \int_0^{2 \pi} d \varphi_{ \bm b }~ \lvert \mathcal{S}_{2C}^{DR} \rvert^2, 
\label{d_sigma_2C^DR}
\end{eqnarray} 
where the integrations run over the absolute value $b$ and the azimuthal angle $\varphi_{\bm b}$ of the impact parameter $\bm b$ and we assume that 
$ b_{min} \gg 1 $ a.u.. The total number of the two-center recombination events is proportional to 
\begin{eqnarray}
\sigma_{2C}^{ DR} = \int d^3 k \, \, \frac{d^3 \sigma_{2C}^{ DR}}{d {\bm k}^3} .
\label{sigma_2C^DR}
\end{eqnarray} 
One should note that since two-center recombination is 
a three-body collision process (incident electron + $A$ + $B$) 
the quantities (\ref{d_sigma_2C^DR}) and (\ref{sigma_2C^DR}) 
strictly speaking are not cross sections (how they are normally defined). 

Although analytical expressions for (\ref{d_sigma_2C^DR}) and 
(\ref{sigma_2C^DR}) can be obtained for bound states $\phi_0$ and 
$\chi_0$ with arbitrary principal and orbital quantum numbers 
(see \cite{Master}), they in general turn out to be quite cumbersome. 
Therefore, in this communication 
we present results only when $\phi_0$ and $\chi_0$ are $s$-states. 

The frequency (energy) spectrum of emitted photons is proportional 
to the 'cross section' 
\begin{eqnarray}
\frac{d \sigma_{2C}^{DR}}{d \omega_k} &=& \frac{ 1 }{ 6 \pi } 
\frac{ \omega_B \omega_k \, r_A^2 r_B^2}{ v^2 b_{min}^2 p^2 } \frac{\Gamma_r^B}{ \frac{(\Gamma_r^B)^2}{4} + \delta^2 } \times \nonumber \\
& & \eta_m^2 \, \bigg\{ \sin^2\vartheta_{\bm p} \, K_1^2 (\eta_m ) + 
 \nonumber \\
& & ~ (1 + \cos^2\vartheta_{\bm p}) \, \eta_m \, K_0 (\eta_m) 
K_1 (\eta_m) \bigg\}. 
\label{en-spectra}
\end{eqnarray}
Here, $p$ and $\vartheta_{\bm p}$ are the absolute value and the polar 
angle, respectively, of the momentum $ \bm p$ of the incident electron, 
and $ \eta_m = | \varepsilon_p - \varepsilon_0 - \omega_k | b_{min}/v$. 
Further, $ r_A = \int_0^{\infty} dr ~ r^3 g_0 (r) g_{p1} (r) $ is the radial matrix element for the transition of the incident electron into 
the ground state $\phi_0$, where $g_{p1}$ and $g_0$ are the radial parts 
of the continuum and bound state, respectively. Similarly, $ r_B = \int_0^{\infty} d \xi ~ \xi^3 h_1^{*} (\xi) h_0 (\xi) $ denotes the radial matrix element for transitions between the ground and excited state of atom $B$ with $h_0$ and $h_1$ being their radial parts. 

The total 'cross section' is obtained by integrating (\ref{en-spectra}) 
over the photon frequency $\omega_k$. In order to perform this integration 
we remark that the right-hand side of (\ref{en-spectra}) contains 
a factor $\Gamma_r^B/(\delta^2 + (\Gamma_r^B)^2/4) $ 
which varies with $\omega_k$ 
much more rapidly than the rest: it has 
a maximum at $ \omega_k = \omega_B = \epsilon_1 - \epsilon_0 $, 
very quickly decreases when the detuning $ | \omega_k - \omega_B |$ increases and is already strongly suppressed 
when the detuning exceeds just several $ \Gamma_r^B $-s 
whereas the other $\omega_k$-dependent factors in (\ref{en-spectra}) 
vary on much broader scales. By exploiting this feature we obtain 
\begin{eqnarray}
\sigma_{2C}^{ DR} &=& \frac{ 1 }{ 3 } \, \frac{ r_B^2 }{ v^2 \, b_{min}^2 } 
\frac{ r_A^2 }{ p^2 } \, 
\tilde{\eta}^2 \bigg\{ \sin^2\vartheta_{\bm p} K_1^2 (\tilde{\eta}) + 
\nonumber \\
& & \qquad \qquad \quad (1 + \cos^2\vartheta_{\bm p}) \, \tilde{\eta} \, 
K_0(\tilde{\eta}) \, K_1(\tilde{\eta}) \bigg\}, \nonumber \\
\label{sigma_2C^DR_1}
\end{eqnarray}
where $\tilde{\eta} = | \varepsilon_p - \varepsilon_0 - \omega_B | b_{min}/v$ and $r_B^2$ can be expressed via the radiative width $\Gamma_r^B$ of atom $B$ according to 
\begin{eqnarray}
 r_B^2 = \frac{9 c^3}{ 4 \omega_B^3 } \Gamma_r^B. 
\label{r_B^2} 
\end{eqnarray}

Taking into account that species $A$ move in a gas of atoms $B$, 
the total decay rate per unit of time for 2CDR per one 
(e$^-$+$A$) pair reads 
\begin{eqnarray}
\mathcal{R}_{2C}^{ DR} &=& \sigma_{2C}^{DR} n_B v 
\nonumber \\ 
& = & 
\frac{ 3 }{ 4 } \, \frac{ n_B }{ v \, b_{min}^2 } \, 
\frac{ c^3 \Gamma_r^B }{ \omega_B^3 } \, 
\frac{ r_A^2 }{ p^2 } \, 
\tilde{\eta}^2 \bigg\{ \sin^2\vartheta_{\bm p} K_1^2(\tilde{\eta}) 
\nonumber \\
&& + (1 + \cos^2\vartheta_{\bm p}) \, \tilde{\eta} \, K_0(\tilde{\eta}) \, K_1 (\tilde{\eta}) \bigg\},
\label{R_2C^DR_0}
\end{eqnarray}
where $n_B$ is the density of atoms $B$. 
The functions $K_n(x)$ ($n=0, 1,..$) diverge at $ x \to 0$ 
and decrease exponentially at $x > 1$ \cite{IStegun}. 
Therefore, in distant low-velocity collisions 
($b_{min} \gg 1$, $v \ll 1$) the most favourable conditions for 2CDR, according to (\ref{R_2C^DR_0}), are realized when the energy of the incident electrons is 
within the small interval centered at $ \varepsilon_{p, r} = \varepsilon_0 + \omega_B $ with the width $ \sim \delta \varepsilon_p \sim v/ b_{min} $. 
Since the quantity $v/ b_{min}$ is typically orders 
of magnitude larger than the natural width $\Gamma_r^B$ we see that the collision strongly 
smears out the 'static' resonance conditions $ \varepsilon_0 + \omega_B - \Gamma_r^B \lesssim \varepsilon_p \lesssim \varepsilon_0 + \omega_B + \Gamma_r^B $ leading to a much broader range of 'quasi-resonance' energies of the incident electron. 

If the incident electrons do not have 
a fixed momentum ${\bm p}$ 
the rate (\ref{R_2C^DR_0}) should be averaged 
over their momentum distribution function $f({\bm p})$. 
This, in general, can be done only numerically. 

However, a simple formula 
for the averaged rate, 
which enables one to establish a 
direct correspondence with the case of 2CDR at 
a fixed distance between $A$ and $B$, 
can be derived if we suppose the following: 
i) the function $f({\bm p})$ 
can be factorized as 
$ f({\bm p}) = f_\varepsilon(\varepsilon_p) \, 
f_\Omega(\Omega_{\bm p}) $;  
ii) the function $f_\varepsilon(\varepsilon_p)$ 
is distributed over an energy range which covers 
the interval of the 'quasi-resonance' energies, 
$ \varepsilon_0 + \omega_B - v/ b_{min} 
\lesssim \varepsilon_p \lesssim 
\varepsilon_0 + \omega_B + v/ b_{min} $, 
and is much broader than this interval 
with $f_\varepsilon(\varepsilon_p)$ noticeably 
varying on a scale much larger than 
$ \delta \varepsilon_p \sim v/ b_{min} $ 
(i.e. within the energy interval essential 
for 2CDR $f_\varepsilon(\varepsilon_p)$ is roughly a constant).  
Then, taking into account that the 'width' of the continuum 
(i.e. the energy range on which the quantity $r_A^2/p^2$ noticeably varies: 
typically $\sim 10$ eV for atoms and $\sim 1$ eV for negative ions) 
is much larger than $ \delta \varepsilon_p \sim v/ b_{min} $, 
we obtain that the averaged rate is approximately given by 
\begin{eqnarray}
\langle \mathcal{R}_{2C}^{ DR} \rangle &=& 
\frac{ 9 \pi^4 }{ 16 } \, \frac{ n_B }{ b_{min}^3 } \, 
\frac{ \Gamma_r^B c^3 }{ \omega_B^3 } \, 
\left( \frac{ r_A^2 }{ p^2 } \right)_{ p = p_r } 
f_\varepsilon(\varepsilon_{p, r}) 
\nonumber \\ 
&& \times \int d\Omega_{\bm p} f_\Omega(\Omega_{\bm p}) 
\left( 1 + \frac{1}{2} \sin^2\vartheta_{\bm p} \right),
\label{R_2C^DR_0-averaged}
\end{eqnarray}
where $ p_r = \sqrt{ 2 \varepsilon_{p, r} } = \sqrt{ 2 (\varepsilon_0 + \omega_B) }$. 
Assuming for simplicity that all electrons are incident 
under the angle $\vartheta_{\bm p} = \pi/2$ and are homogeneously 
distributed over the energy interval $\Delta E$ we get 
\begin{eqnarray}
\langle \mathcal{R}_{2C}^{ DR} \rangle &=& 
\frac{ 3^3 \pi^4 }{ 2^5 } \, 
\frac{ n_B }{ b_{min}^3 } \, 
\frac{ c^3 }{ \omega_B^3 } \, 
\frac{ \Gamma_r^B }{ \Delta E } \, 
\left( \frac{ r_A^2 }{ p^2 } \right)_{ p = p_r }. 
\label{R_2C^DR_0-averaged-simple}
\end{eqnarray}

\subsection{ Single-center radiative recombination } 

Single-center radiative recombination is a very well known process, 
which has been studied for decades with energies of the incident electrons 
ranging from below $1$ eV to relativistic values 
(see e.g. \cite{Hahn}, \cite{Beiersdorfer}, \cite{eik-2007} 
and references therein). 

The (total) rate per unit time for radiative recombination 
of (e$^-$+$A$) pair reads 
\begin{eqnarray} 
\mathcal{R}_{1C}^{RR} = \frac{ 4 \, \pi }{ 3 } \, 
\frac{ \omega_A^3 }{ c^3 } \, \frac{ r_A^2 }{ p^2 }, 
\label{RR-rate-total}
\end{eqnarray}
where $p$, as before, is the momentum of the incident electron, 
$\omega_A = \varepsilon_p - \varepsilon_0 $ is the transition energy 
and $r_A$ is the radial matrix element (which was already defined in the previous subsection). 

If the energy of the incident electrons is not fixed one should 
average the rate (\ref{RR-rate-total}) over their energy distribution. 
Assuming that the width of this distribution is much smaller than 
the energy range on which the quantity $r_A^2/p^2$ noticeably varies 
we obtain that the averaged rate for RR, 
$ \langle \mathcal{R}_{1C}^{RR} \rangle $
simply coincides with $\mathcal{R}_{1C}^{RR}$ 
given by formula (\ref{RR-rate-total}): 
\begin{eqnarray} 
\langle \mathcal{R}_{1C}^{RR} \rangle = \frac{ 4 \, \pi }{ 3 } \, 
\frac{ \omega_A^3 }{ c^3 } \, \frac{ r_A^2 }{ p^2 }.  
\label{RR-rate-total-averaged}
\end{eqnarray} 

\subsection{ 2CDR-to-RR Ratios} 

The relative effectiveness of collisional 2CDR 
and single-center RR can be characterized by the ratios 
\begin{eqnarray}
\mu_{2C, 1C} &=& \frac{ \mathcal{R}_{2C}^{DR} }{ \mathcal{R}_{1C}^{RR} } 
\nonumber \\ 
&=& 
\frac{ 9 }{ 16 \pi } \, \frac{ n_B }{ v \, b_{min}^2 } \, 
\frac{ c^6 \Gamma_r^B }{ \omega_A^3 \, \omega_B^3 } \, 
\tilde{\eta}^2 \bigg\{ \sin^2\vartheta_{\bm p} K_1^2(\tilde{\eta}) 
\nonumber \\
&& + (1 + \cos^2\vartheta_{\bm p}) \, \tilde{\eta} \, K_0(\tilde{\eta}) \, K_1 (\tilde{\eta}) \bigg\} 
\label{ratio-energy-dependent}
\end{eqnarray}
and 
\begin{eqnarray}
\overline{ \mu }_{2C, 1C} & = & 
\frac{ \langle \mathcal{R}_{2C}^{DR} \rangle }{ \langle \mathcal{R}_{1C}^{RR} \rangle} 
\nonumber \\ 
& = & 
\frac{ 3^3 \pi^3 }{ 2^6 } \, \frac{ n_B }{ b_{min}^3 } \, 
\frac{ c^6 \Gamma_r^B }{ \omega_A^3 \, \omega_B^3 } \, 
f_\varepsilon(\varepsilon_{p, r}) 
\nonumber \\ 
&& \times \int d\Omega_{\bm p} f_\Omega(\Omega_{\bm p}) 
\left( 1 + \frac{1}{2} \sin^2\vartheta_{\bm p} \right)
\nonumber \\ 
& = & 
\frac{ 3^4 \pi^3 }{ 2^7 } \, \frac{ n_B }{ b_{min}^3 } \, 
\frac{ c^6 }{ \omega_A^3 \, \omega_B^3 } \, 
\frac{ \Gamma_r^B }{ \Delta E }, 
\label{ratio-averaged}
\end{eqnarray}
where in obtaining the last line of (\ref{ratio-averaged}) 
it was assumed that the electrons are incident under the angle 
$ \vartheta_{\bf p} = \pi/2$ and we set 
$f_\varepsilon(\varepsilon_{p, r}) = 1/\Delta E$. 
 
\subsection{ Collisional 2CDR versus 'static' 2CDR. }

In case of 2CDR occurring at a fixed distance $R_0$ between 
the centers $A$ and $B$ the ratio of this process 
to the single-center radiative recombination 
is given by 
\cite{2CDR}
\begin{eqnarray}
\overline{ \mu }_{2C, 1C}^{static} & \simeq &  
\frac{ c^6 }{ R_0^6 \, \omega_A^3 \, \omega_B^3 } \, 
\frac{ \Gamma_r^B }{ \Delta E }. 
\label{ratio-static-averaged}
\end{eqnarray}

Comparing (\ref{ratio-static-averaged}) with 
(the last line of) (\ref{ratio-averaged}) we see 
that in collisions the role of the fixed 
inter-atomic distance $R_0$ is overtaken by 
$R_{eff} = (b_{min} \, \bar{R})^{1/2}$, 
where $ \overline{R} \approx n_B^{-1/3} $ 
is the average distance between the atoms. 
Thus, the quantity $R_{eff} $ 
plays the role of an effective 
inter-atomic distance in the collisions. 

Since for not very dense gases one has $ b_{min} \ll \overline{R} $ 
we obtain that $ R_{eff} \ll \overline{R} $. Due to a steep dependence 
of the two-center channel on the inter-atomic distance the colliding atoms 
interact mainly in the vicinity of their closest rapprochement 
($R  \sim b$),  which is much less than the averaged distance $ \overline{R} $ between them. 
This explains why the effective distance 
$R_{eff} $ is strongly reduced as compared to the average distance 
$ \overline{R} $. 
Because of the same reason 
the 'electrostatic' form (\ref{2c-ee-int}) of the two-center electron-electron interaction 
may be used provided $ b_{min} \ll c/\omega $. 

\subsection{ Inter-atomic coulombic electron capture }

If we consider a three body collision -- incident electron + $A$ + $B$ -- in the same way as for 2CDR, but now the energy of the incident electron is sufficient to ionize atom $B$, this process is called interatomic coulombic electron capture (ICEC). It was already studied for the 'static' case, in which $A$ and $B$ constitute a bound system \cite{Gokhberg}, 
\cite{sissourat}. 
A detailed consideration of this process in slow atomic collisions is given in \cite{collisionalICEC} and here we only quote our results for the total rate per unit time $\mathcal{R}_{2C}^{EC}$ for ICEC and the ratio 
$\mathcal{\mu}_{1C, 2C}^{EC}$ 
between the total rates for ICEC and single-center RR, 
which are given by  
\begin{eqnarray}
\mathcal{R}_{2C}^{EC} &=& \frac{1}{ 2^{11} \pi} \frac{n_B}{b_{min}^3} ( 5 + \cos^2(\vartheta_{\bm p} ) ) \frac{r_A^2}{p^2} \bigg( \frac{r_B^2}{p_B} \bigg)_{p_B = p_s}
\end{eqnarray}
and
\begin{eqnarray}
\mathcal{\mu}_{1C, 2C}^{EC} &=& \frac{ \mathcal{R}_{2C}^{EC} }{ \mathcal{R}_{1C}^{RR} } \nonumber \\
&=& \frac{ 9 \pi }{ 2^{11} } \bigg( \frac{c}{\omega_A} \bigg)^4 \frac{n_B}{b_{min}^3} ( 5 + \cos^2(\vartheta_{\bm p} ) ) \sigma_{PI}^B(\omega_A). \nonumber \\
\label{ratio-ICEC}
\end{eqnarray}
Here, $r_B$ is the radial matrix element for the bound-continuum transition in atom $B$, $p_B$ the momentum of an electron emitted from $B$, $p_s =  \sqrt{ 2 (\epsilon_0 + \omega_A) }$ and 
$\sigma_{PI}^B$ is the photoionization cross section of atom $B$. 
 
\section{ Results and Discussion }

In order to illustrate our theoretical findings 
here we discuss the relationship between 2CDR 
and single-center RR for a few collision systems. 

Let us first consider the collision system (K$^{+}$ + e$^{-}$) (
$\lvert \varepsilon_0 \rvert \approx 4.34$ eV) -- Be($2s^2$) (atom $B$) in which electron capture by center $A$ is accompanied by excitation of the $2s_{1/2} \rightarrow 2p_{3/2}$ dipole transition in Be ($\omega_B = 5.28$ eV, $\Gamma_r^B = 1.66 \times 10^{-8}$ eV). Using \eqref{ratio-energy-dependent} and choosing $b_{min} = 5 $ a.u., $v = 0.01$ a.u. (corresponding to $2.5$ eV/u), $\varepsilon_p = 0.938$ eV and $\vartheta_{\bm p} = \pi/2$ we obtain that $\mu_{2C, 1C} \geq 1$ if $n_B \gtrsim 1.17 \times 10^{14}$ cm$^{-3}$. An atomic density of $n_B = 10^{15}$ cm$^{-3}$ 
(which is more than four orders of magnitude smaller than the density of air $n_{air} \approx 3 \times 10^{19}$ cm$^{-3}$ under normal conditions) would yield a ratio of $\mu_{2C, 1C} \approx 8.60$.

As a second example let us take the collision system (Cs$^{+}$ + e$^{-}$) ($\lvert \varepsilon_0 \rvert \approx 3.89$ eV) -- Mg($3s^2$) (atom $B$) 
assuming that the dipole transition $3s \rightarrow 3p$ in Mg ($\omega_B = 4.35 $ eV, $\Gamma_r^B = 3.35 \times 10^{-7}$ eV) is involved. 
Applying \eqref{ratio-energy-dependent} with 
$b_{min} = 5$ a.u., $v = 0.01$ a.u., $\varepsilon_p = 0.458$ eV and $\vartheta_{\bm p} = \pi/2$ we obtain $\mu_{2C, 1C} \geq 1$ if $n_B \gtrsim 4.90 \times 10^{13}$ cm$^{-3}$. Choosing an atomic density of $n_B = 10^{15}$ cm$^{-3}$ would lead to $\mu_{2C, 1C} \approx 20.4$ (see Fig. 2). 

Finally, we consider the collision system (Li$^{+}$ + e$^{-}$) 
($\lvert \varepsilon_0 \rvert \approx 5.39$ eV) -- H($1s$) (atom $B$) 
in which electron capture by Li is assisted by the dipole transition $1s \rightarrow 2p$ in H ($\omega_B = 10.2$ eV, $\Gamma_r^B = 7.44 \times 10^{-6}$ eV). Employing \eqref{ratio-energy-dependent} with $b_{min} = 5 $ a.u., $v = 0.01$ a.u., $\varepsilon_p = 4.81$ eV and $\vartheta_{\bm p} = \pi/2$, we obtain that $\mu_{2C, 1C} \geq 1$ if $n_B \gtrsim 1.06 \times 10^{15}$ cm$^{-3}$. An atomic density of $n_B = 3 \times 10^{15}$ cm$^{-3}$ yields $\mu_{2C, 1C} \approx 2.85 $. 

Thus, the 2CDR channel can dominate single-center RR of a free electron with atomic center $A$ for relatively low densities of atoms $B$ (as compared to $n_{air}$). 
One reason for the good performance of the 2CDR channel is that $R_{eff} \ll \overline{R}$. For example, using $b_{min} = 5$ a.u. and $n_B = 10^{15}$ cm$^{-3}$, we obtain $ R_{eff} \approx 5 $ nm $ \ll \overline{R} \approx 100$ nm. 

\begin{figure}[h!]
\vspace{-0.25cm}
\begin{center}
\includegraphics[width=0.5\textwidth]{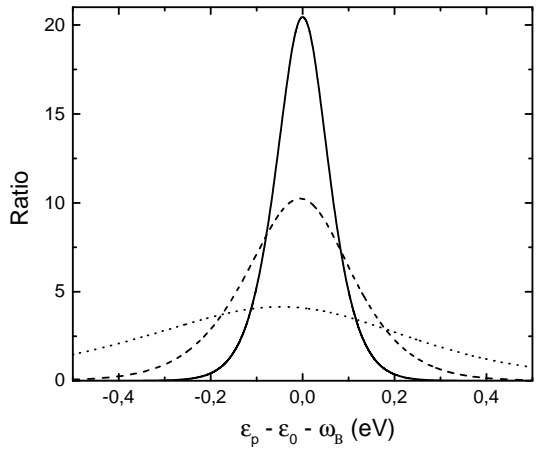}
\end{center}
\vspace{-0.5cm}
\caption{ The ratio (\ref{ratio-energy-dependent}) as a function 
of $ \Delta \varepsilon_p = \varepsilon_p - \varepsilon_0 - \omega_B $ 
at $\vartheta_{\bm p} = \pi/2$  for collision velocities $v = 0.01$ a.u. (solid), $v = 0.02$ a.u. (dashed) and $v = 0.05$ a.u. (dotted) 
for the collision system (Cs$^{+}$ + e$^{-}$) -- Mg($3s^2$).
$b_{min} = 5 $ a.u., $n_B = 10^{15}$ cm$^{-3}$. }
\label{figure2}
\end{figure} 

For more insight, in Fig. \ref{figure2} we show the ratio $\mu_{2C, 1C}$ 
given by \eqref{ratio-energy-dependent} as a function of 
the detuning $ \Delta \varepsilon_p = \varepsilon_p - \varepsilon_0 - \omega_B $ from the 'resonance' energy of the incident electron  
at a fixed $\vartheta_{\bm p} = \pi/2$. Since the shape of 
$\mu_{2C, 1C}$ turns out to be very similar for all the collision systems considered above, in Fig. 2 it is presented just for one of them, 
(Cs$^{+}$ + e$^{-}$) -- Mg($3s^2$), for impact velocities 
ranging from $0.01$ a.u. ($2.5$ eV/u) to $0.05$ a.u. ($62.5$ eV/u). 
It follows from the figure that the function $\mu_{2C, 1C}$ reaches 
a maximum at the position of the resonance, 
$ \varepsilon_{p,r} = \varepsilon_0 + \omega_B $, 
and is roughly symmetric with respect to this point. 
The maximum is rather broad: its width is caused by 
the relative motion of centers $A$ and $B$ and even 
for the lowest velocity considered in Fig. \ref{figure2} 
it exceeds the corresponding radiative width 
$ \Gamma_r^B $ ($ \Gamma_r^B \sim 10^{-7}$ eV) 
by many orders of magnitude.    

In general, the total width of the excited state 
of atom $B$ interacting with center $A$ is determined by the sum of its radiative width 
$ \Gamma_r^B $ and the width $ \Gamma_{2c-A} $ due to 
the (two-center) Auger decay of this state caused by two-center electron-electron interaction. 
When the distance between $A$ and $B$ becomes not very large ($ \stackrel{<}{\sim } 10 $ a.u.) 
the Auger width $ \Gamma_{2c-A} $ begins to exceed the radiative width 
$ \Gamma_r^B $ \cite{2CPI}. 
The width $ \Gamma_{2c-A} $ is, nevertheless, ignored in our description  because of the following. Even at the low impact velocities $v$ 
considered here the time $T$ which $A$ and $B$ spend in the collision at those distances, where the Auger width $ \Gamma_{2c-A} $ becomes close or even exceeds the radiative width $ \Gamma_r^B $, is so short that $ \Gamma_{2c-A} \, T \ll 1$, i.e. the Auger decay simply does not have enough time to unveil itself in the collision. 

At this point one more remark can be appropriate. 
In our treatment of collisional 2CDR we use bound states of free (non-interacting) centers $A$ and $B$. 
In distant collisions, which are considered here, the interaction between them is quite weak. Nevertheless, as estimates show, even in such collisions this interaction may influence these states shifting, their energies by noticeable amounts. Since the latter ones can be much larger than the radiative and Auger widths 
of $B$ the neglect of them would clearly 
be unjustified for considering 2CDR in the 'static' situation 
in which a very fine tuning (within $ \Gamma_r^B $ or $ \Gamma_r^B  +  \Gamma_{2c-A} $) of the transition energies on both centers 
is necessary in order to reach the highest possible effectiveness of the two-center process \cite{2CDR}. 
However, in case of collisional 2CDR the relative motion so strongly broadens the resonance (see Fig. 2) that the neglect of 
the energy shifts is not expected to have a substantial impact 
on the result.        

\vspace{0.15cm} 

It is of interest to compare collisional 2CDR 
with collisional 2CPI which was studied very recently in \cite{2CPI_1}. 
In the 'static' situation, where the two centers constitute a bound system, 
both 2CDR and 2CPI show about the same effectiveness 
compared to the single-center radiative recombination and photo ionization, 
respectively. It turns out, however, that in collisions 2CDR 
becomes substantially less effective compared to the 2CPI 
(which at the first glance might seem unexpected  
since these processes can be thought of as inverse of each other). 

The collision influences the processes of two-center recombination 
and photo ionization in two main ways. First, in collisions 
the effective distance between $A$ and $B$ is greatly increased compared 
to the inter-atomic distance in the 'static' case. This equally impacts 
both collisional 2CDR and collisional 2CPI making them less effective 
than their 'static' counterparts. Second, the relative motion of the centers $A$ and $B$  
effectively broadens their internal transition energies (as they are 'viewed' 
by the collision partner) that in general diminishes the role 
of resonances (both 2CDR and 2CPI are resonant processes). 

However, since in the collisional 2CPI the source of photons 
is at rest with respect to resonating atoms $B$ \cite{2CPI_1},  
the relative motion of centers $A$ and $B$ does not affect 
the resonant character of the first step of this process -- 
the interaction between $B$ and the external laser field: 
like in the 'static' case, atom $B$ acts as a very efficient 'antenna' 
absorbing energy from the laser field and transferring it to 
the subsystem (e$^-$+$A$). Although from the 'point of view' of 
the latter the transfer involves a rather broad range of energies,  
this does not affect its effectiveness since 
transitions in (e$^-$+$A$) are between bound and continuum states 
and, thus, are not resonant. 

In contrast, in the collisional 2CDR its first step -- 
the energy transfer between the internal states of 
the subsystems (e$^-$+$A$) and $B$ -- is strongly affected by 
the relative motion. From the 'point of view' of atom $B$ 
this motion broadens the energy of electron transitions 
in (e$^-$+$A$) and even at low collision velocities this broadening 
is much larger than the natural width of the excited state of $B$. 
As a result, there is a very low probability that,  
for a given change $\varepsilon_p - \varepsilon_0$ in the internal energies 
of (e$^-$+$A$), the corresponding energy transfer $ \omega_{tr} $ to $B$ will fit into the resonance conditions $ \omega_B - \Gamma_r^B \stackrel{<}{\sim} \omega_{tr} \stackrel{<}{\sim} \omega_B + \Gamma_r^B$ for the spontaneous radiative decay of $B$.  
That is why the two-center process studied in the present paper 
is less effective compared to collisional two-center photo ionization. 
   

\vspace{0.25cm} 

Let us now very briefly consider the correspondence 
between single-center RR and the ICEC in collisions.   
In order to compare these processes we need incident electron energies which are higher than in the 2CDR since now atom $B$ is ionized. 
Taking the collision system (Cs$^{+}$ + e$^{-}$) -- Mg($3s^2$) and  
choosing $ b_{min} = 5 $ a.u., $ v = 0.01 $ a.u., 
$\varepsilon_p = 3.81$ eV we obtain that the transition frequency for the electron capture, Cs$^{+}$ + e$^{-}$ $\rightarrow$ Cs, is 
$\omega_A = 7.70$ eV which is slightly above the ionization threshold 
for Mg($3s^2$) ($ | \epsilon_0 | = 7.65$ eV). Then, we can use the experimental photo ionization cross section \cite{Ditchburn} to get 
$\sigma_{PI}^{\text{Mg($3s^2$)}} (7.70 \text{ eV}) \approx 1.2 \times 10^{-18}$ cm$^2$. For $\vartheta_{\bm p} = \pi/2$ we obtain 
$\mu_{2C, 1C} \geq 1$ if $n_B \gtrsim 5.2 \times 10^{18}$ cm$^{-3}$. 
Thus, compared to the 2CDR, an increase in 
the atomic density $n_B$ by more than four orders is necessary 
to make the ICEC channel comparable in strength  
to the corresponding single-center radiative recombination.  

Such a large difference between the effectiveness of the 2CDR 
and the ICEC is caused mainly by two reasons. 
One of them (minor) is larger transition frequencies involved in the ICEC but the main reason is that a dipole allowed transition from the ground state to an excited bound state, which is characteristic for the 2CDR, 
may be comparable (or even exceed) in its strength transitions from the ground state to the whole (single-electron) continuum (we remark that 
the ICEC involves just a tiny fraction of this continuum).  

\section{ Conclusions } 

We have considered two-center dielectronic recombination occurring 
in slow atomic collisions. Our consideration was based on the semi-classical approximation and the first order of perturbation theory in 
the inter-atomic interaction. Only contributions to this process from relatively large impact parameters were taken into account which means 
that the present results should in fact be viewed as yielding 
a lower boundary for the effectiveness of this process in the collisions. 

We have shown that two-center dielectronic recombination, in which 
the capture of an incident free electron by center $A$ is driven by 
dynamic two-center electron-electron correlations involving 'quasi-resonant' dipole-allowed bound-bound transitions in center $B$, 
can outperform the direct single-center process of radiative recombination also in collisions, provided the density of atoms $B$ is not too low. 
Thus, the 2CDR can 'survive' even in collisions where 
the mean distance between $A$ and $B$ exceeds by orders of magnitude 
the typical size of a bound $A$-$B$ system. 

Compared to the process of collisional two-center photo ionization 
\cite{2CPI_1} the process considered in the present paper is more depreciated by the relative motion of the colliding centers. This motion affects quite differently the resonance conditions on which both these processes heavily rely: while these conditions are essentially 
not influenced by this motion in case of collisional 2CPI, it does 
wash out the resonant character 
of collisional 2CDR. 

\section*{Acknowledgement} 

This work was funded in part by the Deutsche Forschungsgemeinschaft 
(DFG, German Research Foundation) under Grant No 349581371 
(MU 3149/4-1 and VO 1278/4-1).  


\end{document}